\documentclass[12pt]{article}
\usepackage[english]{babel}
\usepackage{feynmf}
\usepackage{amssymb}
\def\nn{\nonumber}
\def\mi0{\mu_0}
\def\ba{\begin{eqnarray}}
\def\ea{\end{eqnarray}}
\def\vB{\vec{B}}
\def\vE{\vec{E}}
\def\cB{{\cal B}}
\def\vcE{\vec{\cal E}}
\def\vA{\vec{A}}
\begin{document}
\title{{On Massive Vector Bosons and Abelian Magnetic Monopoles in $D=(3+1)$: a Possible Way to Quantise the Topological Mass Parameter}\thanks{Dedicated to our Master Prof. Jos\'e Leite Lopes on the occasion of his 80$^{\mbox{th}}$ birthday.\ Earlier versions of this manuscript appeared in Ref.[18,19].}\author{ Winder A. Moura-Melo\thanks{Email:winder@cbpf.br.}, N. Panza and J.A. Helay\"el-Neto\thanks{Also at Universidade Cat\'olica de Petrop\'olis (UCP)
.\ Email: helayel@cbpf.br.}\\ \\ Centro Brasileiro de Pesquisas F\'{\i}sicas \\ Rua Xavier Sigaud 150 - Urca\\ 22290-180 - Rio de Janeiro, RJ - Brasil.}}
\date{}  \maketitle 
\begin{abstract}
An Abelian gauge model, with vector and  2-form potential fields linked by a topological mass term, is shown to exhibit Dirac-like magnetic monopoles in the presence of a matter background.\ In addition, considering a `non-minimal coupling' between the fermions and the tensor fields, we obtain a generalized quantization condition that involves, among others, the topological mass parameter.\ Also, it is explicitly shown that 1\_loop (finite) corrections do not shift the value of such a mass parameter.
\end{abstract} 
\section*{Introduction}
Magnetic monopoles were firstly proposed by Dirac \cite{Dirac1} in the framework of Classical Electrodynamics with the main aim to provide a physical explanation of why the electric charges appear only as integer multiples of the elementary one (electron or proton charge, denoted by $e$).\ Indeed, Dirac obtained that {\em `` if there exists one quantum magnetic pole in Nature, $g_o$, interacting with electric charges, then Quantum Mechanics demands the quantization of the latter according to}:\footnote{We are using Lorentz-Heaviside's units for Electrodynamics.\ In his original paper Dirac \cite{Dirac1} used Gaussian ones.} 
\ba
q g_o=2\pi \hbar c \quad, {\rm{\mbox with}}\quad q=ne \quad  n \quad{\rm{\mbox integer. ''}} \label{dcq}
\ea
Among other features, his work pointed out the relation between gauge invariance and the singular structure of gauge potentials, the non-physical string (see also Ref.\cite{GOlive}, section 2.5 and Ref.\cite{Giamb}).\\

\begin{sloppypar}In general, these objects are `{\em put in by hand}' in Electrodynamics-type models (Maxwell, Proca, etc.) by breaking the Bianchi's identity of the $A_\mu$-sector (so, circumventing the Poincar\'{e}'s lemma on differential forms).\ Their presence restore the {\em duality} between the electric and magnetic sectors, lost after the introduction of the electric current.\ Therefore, Dirac's monopoles render Electrodynamics more symmetric, and the $U(1)$-gauge group {\em compact}: the Abelian and unitary operator $S$ which implements the gauge transformations becomes {\em single-valued}.\ In particular, this aspect is crucial for non-Abelian theories which have their vacuum symmetry broken by scalar fields (Higgs' mechanism).\ In these cases, if the original non-Abelian gauge group of the vacuum is broken to $U(1)$-compact group, then the classical  dynamical equations yield (static) soliton solutions carrying (Abelian) magnetic charge (at large distances, looking as Dirac's monopoles).\ This was firstly shown by 't Hooft \cite{tHooft} and Polyakov \cite{Polyakov}, dealing with the Georgi-Glashow's \cite{GG} model; see Ref.\cite{GOlive}, Sections 5 and 6, for the extension to arbitrary {\em simply-connected} gauge groups (see also the references listed in \cite{extension}).\ Recently, it was shown that $N=2$-supersymmetric Yang-Mills theories present {\em monopole condensation}, which seems to be essential for understanding quark confinement \cite{SW}.\ Eventually, if such non-Abelian gauge theories (supersymmetric or not) are correct, then their magnetic monopoles {\em must exist}.\footnote{Nevertheless, the observation of such objects is deeply jeopardized by their huge masses.\ For example, for $SU(5)$- gauge group these masses are of the order of 10$^{\mbox{16}}$Gev, increasing as the gauge group is enlarged.\ See, for example, Ref. \cite{Giacomelli}.}\\ \end{sloppypar}

\begin{sloppypar}There are some similarities and differences between 't Hooft-Polyakov' and Dirac's monopoles.\ Here, we wish to pay attention to one of these differences: while the first type coexists with massive vector boson (the masses of both being given by the scalar fields, after the spontaneous symmetry breaking) the same does not happen to the second one.\ In fact, it seems that for Abelian theories (defined on Minkowski's flat space-time), Dirac's monopoles {\em can appear only if} the vector boson is {\em massless} \cite{YJ}-\cite{Dattoli}.\ This has been shown in several works to be true for the Proca's model (the simplest finite-range extension of Maxwell's theory, where the  boson mass stems from explicity breaking the gauge symmetry) and, in addition, some attempts have been made to bypass this impossibility, by considering pairs of monopoles (with opposite charges) joined by a Dirac' string \cite{Ahrens}, or even the presence of a `{\em massive tachyon}' as being the {\em superluminal counterpartner} of the `physical massive photon' \cite{Dattoli}. \\ \end{sloppypar}
It is precisely on this subject that lies our motivation for this work: are there any physical arguments that rule out the coexistence of both massive vector bosons and Dirac monopoles within an Abelian model defined on flat Minkowski space-time? Would such an impossibility arise from the structure of a particular theory or from the specific {\em mechanism} for gauge boson mass generation?\\ 
At the attempt of taking some glance on this question, we shall study a particular model, within which two Abelian factors (a vector and a 2-form gauge potentials) are linked by a {\em topological mass} term, giving us a massive vector boson as its particle physical content, Ref.\cite{CS,KR}.\\

We should stress that, we are not presenting here a general proof for the question raised above; rather, our purpose is to provide one more explicit example of a theory in which Dirac-like monopoles do not show up while the intermediate gauge boson is massive.\ The particular mechanism for gauge-field mass generation does not seem to be relevant for the suppression of the monopoles: once the gauge-field propagator develops a non-trivial pole, Dirac monopoles are ruled out (we shall come to this matter throughout our paper).\\
To conclude the presentation of the arguments that motivate our work, we should draw the attention to a peculiar feature: the monopole appearing in the present model (see Section 1) is such that the gauge-field mass parameter enters the charge quantization relation, as it will become clear at the end of our paper.\\
This paper is outlined as follows: in Section 1, we start by presenting the model as well as some of its basic characteristics.\ In Section 2, we show that the model under consideration does not admit, consistently, the coexistence of both Dirac's monopoles and massive vector boson, unless we take a special {\em ansatz} for the current, previously incorporated in the model interacting with $A_\mu$ gauge field.\ We start Section 3 by allowing an `extra-coupling' between the fermionic current and the tensorial gauge sector, by means of a gauge and Lorentz-invariant term.\ In addition, it is shown that if the current {\em ansatz} is implemented, we get a {\em generalized} quantization condition, which contains, among others, the mass parameter.\ This section is closed with a discussion on the no-shift of the topological mass parameter by (finite) 1\_loop contributions.\ The relevant Feynman's graph and its result are presented in the Appendix.\ Finally, we conclude the paper by making a brief discussion about the results and some possible consequences of them.   

\section{The model and some basic aspects}
The Cremmer-Scherk-Kalb-Ramond (CSKR) model \cite{CS,KR} in
the absence of matter fields reads:
\ba
{\cal L}_1 = -\frac1 4 F_{\mu\nu}F^{\mu\nu} +\frac1 6
G_{\mu\nu\kappa}G^{\mu\nu\kappa} +\mi0
\epsilon_{\mu\nu\kappa\lambda}A^\mu \partial^\nu H^{\kappa
\lambda}\, , \label{1}
\ea
with the definitions for the field strengths:
\ba
F_{\mu\nu}=\partial_\mu A_\nu -\partial_\nu A_\mu \qquad
{\rm{\mbox  and}}\qquad
G_{\mu\nu\kappa}=\partial_\mu H_{\nu\kappa} +\partial_\nu
H_{\kappa\mu} +\partial_\kappa H_{\mu\nu}\, , \label{2}
\ea
$H_{\mu\nu}=-H_{\nu\mu}$.\ Here, we are using Minkowski 
metric $diag(\eta_{\mu\nu})=(+,-,-,-)$ and $\epsilon^
{0123}=+1=-\epsilon_{0123}$ for the four-dimensional
Levi-Civita symbol; greek indices run $0,\ldots 3$;
latin characters go from 1 to 3.

As it can be easily checked, the action $S_1=\int dx^4
{\cal L}_1$ is invariant under the independent local Abelian gauge
transformations:
\ba
& & A_\mu (x) \stackrel{U(1)_Lambda}{\longmapsto} A_\mu
'(x)=A_\mu (x) -\partial_\mu \Lambda(x) \, , \label{4} \\
& & H_{\mu\nu}(x) \stackrel{U(1)_\xi}{\longmapsto}
H_{\mu\nu}'(x)=H_{\mu\nu}(x) +\partial_\mu \xi_\nu(x)
-\partial_\nu \xi_\mu(x) \, , \label{5}
\ea
provided that we assume that the parameters $\Lambda$ and
$\xi_\mu$ vanish at infinity.

From (\ref{1}), there follow the field equations:
\ba
& & \partial_\mu F^{\mu\nu}=-\mi0\epsilon^{\nu\kappa
\alpha\beta}\partial_\kappa H_{\alpha\beta}=-\frac
{\mi0}{3}\epsilon^{\nu\kappa\alpha\beta} G_{\kappa
\alpha\beta}\, , \label{6} \\
& & \partial_\mu G^{\mu\nu\kappa}=
+\mi0\epsilon^{\nu\kappa\alpha\beta}\partial_\alpha
A_\beta=+\frac{\mi0}{2}\epsilon^{\nu\kappa\alpha\beta}
F_{\alpha\beta} \, ,\label{7}
\ea
and, from the antisymmetric property of the field strenghts,
we get the Bianchi's identities (geometrical equations):
\ba
\partial_\mu\tilde{F}^{\mu\nu}=0 \qquad {\rm{\mbox  and}}
\qquad \partial_\mu\tilde{G}^\mu=0\, ,\label {8}   
\ea
with: $\tilde{F}_{\mu\nu}=\frac1 2 \epsilon_{\mu\nu\kappa
\lambda}F^{\kappa\lambda} \,{\rm{\mbox  and}} \, \tilde{G}
_\mu =\frac1 6 \epsilon_{\mu\nu\kappa\lambda}G^{\nu\kappa
\lambda}$ defining the dual tensors.\\

The linking term between the gauge fields is {\em topological} because it does not contribute to the gauge-invariant energy-momentum tensor (and so, carrying no energy and propagating no physical degrees of freedom), what is obvious since it requires no metric for its definition (like the Chern-Simons term in 3 dimensions).\ On the other hand, one sees that one gauge field (or more precisely, its field strength) provides a {\em current} for another, and vice-versa, having these currents came about from the topological term.\\ 

The spectrum of the model is the following: if we take $\mi0=0$ (free Lagrangean), $A_\mu$
describes a massless vectorial boson and $H_{\mu\nu}$
behaves as a massless scalar field.\ Therefore, we have 3
degrees of freedom (on-shell).\ In the other case
($\mi0\neq0$), we have a massive vector boson
(with mass $M^2=+2\mi0^2$).\ Here, this particle can be
described by $A_\mu$ as well as by $H_{\mu\nu}$.\ Thus,
in both cases, the model has 3 on-shell degrees of freedom,
what is physically convincent, because the topological term
introduces no additional ones, as we said earlier.\ In fact, it provides a mass generating mechanism, that
was called {\em topological dynamic symmetry breaking} by
Cremmer and Scherk \cite{CS}.\ Kalb and Ramond \cite{KR}
studied it in the context of classical interaction of strings
in dual models.

Moreover, it has been shown that the model is unitary and
renormalizable (in the presence of fermions interacting with
the $A_\mu$ gauge field; the model presented in section 2,
equation (\ref{10})), and also that its mass generating
mechanism is different (at quantum level) from the Higgs when
this is added to the Maxwell theory \cite{ABL}.\ Among others
features, the vacua funtional for the model was obtained by
Amorim and Barcelos-Neto \cite{ABN}.\\
\section{The matter background and the Dirac-type monopole
configuration in the model}
Here, we shall show that, at a na\"{\i}ve step, Dirac's monopoles cannot appear within the CSKR-model.\ Nevertheless, situation can be changed (at low momentum limit) if we introduce matter current in the model satisfying a peculiar relation.
We start by introducing classical configurations of Dirac's magnetic monopole in the CSKR-model.\ This is done by `breaking' the Bianchi's identity for the $A_\mu$-sector\cite{Dirac1,GOlive}, say:\footnote{We shall use the expressions {\em electric} and {\em magnetic} for the $A_\mu$ sector, by its analogy with Maxwell's Electrodynamics.}
\ba
\partial_\mu\tilde{F}^{\mu\nu}=0 \, \stackrel{monopole}
{\longmapsto}\, \partial_\mu\tilde{F}^{\mu\nu}=\chi^\nu\, , 
\label{15}
\ea
where the conserved magnetic 4-current is given by:
$\chi^\mu=(\chi^0,\vec{\chi})$.

For our purposes, should be more convenient to work with the field
equations in vector notation.\ So, we define:
\ba
A^{\mu}\equiv(\Phi,+\vA) \qquad H_{\mu\nu}=\left\{ \begin{array}
{l}H_{0i}\equiv (+\vec{a})_{i} \\H_{ij} \equiv -\epsilon_{ijk}
(\vec{\varphi})_{k}, 
\end{array} \right. \, , \label{17}
\ea
and the field strengths as:
\ba
 F_{\mu\nu}=\left\{\begin{array}{l}F_{0i}\equiv +(\vE)_{i} \\ 
F_{ij}\equiv  -\epsilon_{ijk}(\vB)_{k} \end{array} \right. \qquad  G_{\mu\nu\kappa}=\left\{\begin{array}{l}G_{0ij} \equiv -\epsilon_
{ijk}(\vec{\cal E})_{k}\\ 
G_{ijk}\equiv +\epsilon_{ijk}\, {\cal B} \end{array} \right. \, 
,\label{18}
\ea
which give us: $ \tilde{G}^\mu=(\cB,+\vcE)$. 

Now, the set of equations (\ref{6},\ref{7},\ref{15}) and the identity $\partial_\mu\tilde{G}^\mu=0$, describing a static and point-like magnetic monopole ($\chi^0=+g\delta^3(x);\, \vec{\chi}=0 $ and the static limit for the fields) take the forms:
\ba
& & \nabla\wedge\vB(\vec{r})= {-2\mi0\vec{\cal E}(\vec{r})}  
,\qquad\nabla\cdot\vE(\vec{r})={-2\mi0{\cal B}
(\vec{r})} \label{19}\\
& & \nabla\wedge\vec{\cal E}(\vec{r})= {+\mi0\vB(\vec{r})}  
,\qquad\nabla\cdot\vec{\cal E}(\vec{r})={0} \\ 
& & \nabla\cdot\vB(\vec{r})= \chi^{0}(\vec{r})= g\delta^{3}
(\vec{r}),\qquad \nabla\wedge \vE(\vec{r})=
 {0}  \\
& & \nabla {\cal B}(\vec{r})={-\mi0\vE(\vec{r})} \label{20} 
\ea 
\begin{sloppypar} 
Now, to study the self-consistency of the above equations, we
split them into two sets: one involving the $\cB$ and
$\vE$ fields, and the another with $\vB$ and $\vcE$ vectors
.\ For the first set, it is easy to find good solutions 
\cite{YJ,tese}:
\ba
\vE(\vec{r})= \frac{\vE_{0}}{4\pi}exp(-\sqrt{2}\mi0|\vec{r}|) \quad  {\rm{\mbox and}} \quad {\cal B}(\vec{r})= \frac{{\cal B}_0}
{4\pi}exp(-\sqrt{2}\mi0|\vec{r}|),
\ea
with $\sqrt{2}\mi0{\cal B}_{0}\hat{r}=+\vE_{0}$.\ Nevertheless, for the other set we have troubles: the monopole-like solution that comes from: $\vB(\vec{r})=
+g\vec{r}/4\pi r^3\equiv \vB^D(\vec{r})$ is
inconsistent with $\nabla\wedge\vB(\vec{r})= {-2\mi0\vec{\cal E}
(\vec{r})}$ ($\neq 0$, a priori).\ Even here, we may search for
a more general solution for $\vec{B}$, say $\vec{B}(\vec{r})=
\vB^D(\vec{r}) +\vec{B}'(\vec{r})$ (and similar forms to
$\vec{A}$ and $\vec{\cal E}$ \cite{YJ,tese}) with $\vec{B}'$ given by:$$\vB'(\vec{r})= {\nabla\wedge\frac{\vec{\cal E}'(\vec{r})}{\mi0}=+\frac{2\mi0}{4\pi} \int d^{3}\vec{r'}(1+\sqrt{2}\mi0{R})\, \frac{exp(-\sqrt{2}\mi0{R})}{R^{3}}
\, (\vec{\cal E}^{D}(\vec{r'})\wedge\vec{R})},$$whith $\vec{R}\equiv (\vec{r}-\vec{r'})$.\ Unfortunately, these new solutions prevent us from obtaining a {\em conserved angular momentum operator}, ${\cal J}$, and so from quantise the system of an electrically charged particle placed into this magnetic field\footnote{This point is not so obvious.\ The arguments which lead us to this result are presented in Ref.\cite{YJ}, and are based upon $SU(2)$ algebra analysis.}  (at the non-relativistic limit), whose Lagrangean is $L_p=\frac12 m\dot{r}^2 +q\vec{A}\cdot\dot{\vec{r}}$, with $\vec{A}=\vec{A}^D +\vec{A}'$.\\ Alternatively, based on the Wu-Yang's approach \cite{WuYang}, one can demonstrate the non-existence of a Abelian and unitary operator $S$ which would relate two functions $A^a _\mu$ and $A^b _\mu$, in a overlapping region around the monopole, by a gauge transformation (this is worked out in Ref.\cite{tese}).\ Consequently, at this first stage, the CSKR-model {\em is not } compatible with Dirac's monopoles and this comes about due the massive character of the vector boson.\ In other words, the mass parameter prevents the magnetic field created by the monopole from being spherically symmetric and this, in turn, leads us to the troubles discussed above. \end{sloppypar}
Let us carry on our work and take the CSKR-model with matter fields (say, fermionic).\ The Lagrangean reads:
\ba
{\cal L}_1 \, \stackrel{matter}{\longmapsto} \, {\cal L}_2=
{\cal L}_1 +\overline{\psi}(x)\left(\imath D_\mu\gamma^\mu
-m_f\right)\psi(x)\, ,\label{10}
\ea
with ${\cal L}_1$ already defined in (\ref{1}) and $D_\mu \psi(x)\equiv (\partial_\mu +\imath e A_\mu)\psi(x)$.
\begin{sloppypar}
It is easy to see that $S_2$ is $U(1)_{A\mu}\otimes U(1)_{H\mu\nu}
$-invariant, provided that the fermionic fields transform in the usual way:
$\psi(x)\mapsto \psi'(x)=e^{+\imath e\Lambda(x)}\psi(x)$ and $ \overline{\psi}(x)\mapsto\overline{\psi'}(x)=e^{-\imath e \Lambda(x)}\overline{\psi}(x)$.\end{sloppypar}

From ${\cal L}_2$, the dynamical equations for the
fermions follow:
\ba
(\imath D_\mu\gamma^\mu -m_f)\psi(x)=0 \qquad {\rm{\mbox  and}}
\qquad \overline{\psi}(x)(\imath\stackrel{\leftarrow}{\partial}
_\mu\gamma^\mu +eA_\mu\gamma^\mu +m_f)=0 \, . \label{13}
\ea

Analogously, for the gauge fields, we obtain their dynamical
equations:
\ba
& & \partial_{\mu}F^{\mu\nu}=-\mi0\epsilon^{\nu\kappa\lambda\rho}
\partial_{\kappa}H_{\lambda\rho} + eJ^{\nu}=-2\mi0 \tilde{G}^\nu
+eJ^\nu \label{14a} \, ,\\
& & \partial_\mu G^{\mu\nu\kappa}=+\mi0\epsilon^{\nu\kappa
\lambda\rho}
\partial_{\lambda}A_{\rho}=+\mi0\tilde{F}^{\nu\kappa} \, ,
 \label{14b}
\ea
and also the Bianchi's identities (\ref{8}).\ Here, the conserved
fermionic 4-current is defined by: $J^\mu\equiv\overline{\psi}
\gamma^\mu\psi=(\rho,\vec{J})$.
 
Now, by introducing static and point-like monopole and taking the equations describing it with fermionic 4-current, we get:
\ba
& & \nabla\wedge\vB(\vec{r})= {+ e\vec{J}(\vec{r})-2\mi0\vec{\cal E}
(\vec{r})} ,\qquad\nabla\cdot\vE(\vec{r})={ e\rho
-2\mi0{\cal B}(\vec{r})} \label{35} \\
& & \nabla\wedge\vec{\cal E}(\vec{r})= {+\mi0\vB(\vec{r})}
,\qquad \qquad\nabla\cdot\vec{\cal E}(\vec{r})= {0}\\
& & \nabla\cdot\vB(\vec{r})= {+\chi^{0}=+ g\delta^{3}(\vec{r})}
,\qquad\nabla\wedge\vE(\vec{r})={0} \\
& &  \nabla{\cal B}(\vec{r})={-\mi0\vE(\vec{r})}\, .\label{38}  
\ea
It is clear that, the presence of this current in the above equations leads us to describe {\em another type} of magnetic monopoles, different of those Dirac's ones.\ This difference will be later clarified. 

Let us study the self-consistency of these equations: again, for
the set of $\vec{E}$ and ${\cal B}$ fields it is easy to obtain
well-behaved solutions:$${\cal B}(\vec{r})= -\frac{e\mi0}{4\pi} \, \frac{exp(-\sqrt{2}\mi0 |\vec{r}|)}{|\vec{r}|} \quad , \quad \vE(\vec{r})= \frac{e}{4\pi} \,(r-\sqrt2\mi0 r^2) \,\frac{exp(-\sqrt{2}\mi0|\vec{r}|)}{|\vec{r}|^3}\hat{r}$$Now, to solve the former
problem presented by the another set, we look for
${\cal L}_2$ at low momentum\footnote{Noticing the correspondence: $\imath\partial_\mu \leftrightarrow p_\mu$, we take the low momentum limit by taking $p^2 \ll p$ and write ${\cal L}_2$ up to terms proportional to $p$ (or better $\partial$).\ In words, we consider the kinetical terms {\em small} as compared with others.}
\ba
{\cal L}_2 \stackrel{p\to 0}{\longrightarrow}{\cal L}_{p\to 0}\,
\approx\, +\mi0 \epsilon^{\mu\nu\kappa\lambda}A_\mu\partial_\nu
H_{\kappa\lambda} -e J^\mu A_\mu +({\rm{\mbox  fermionic\, mass \, term}}) \, ,\label{40}
\ea
(` $\approx$ ' stands for {\em approximately to}).\ Here, by calculating the
field equation for $A_\mu$, we get:
\ba
eJ_\mu=+2\mi0\tilde{G}_\mu\, , \label{41}
\ea
Here, we are dealing essentially with the non-relativistic limit (low momentum) of a physical system (particle into a external magnetic field); therefore, it is physically acceptable to take the following {\em ansatz}:\footnote{Let us remind the London's {\em ansatz} for superconductivity: $j_{\mu}=\kappa A_{\mu}$.\ Despite of the nature of the fields ($A_\mu$ is a gauge field and $\tilde{G}_\mu$ a gauge-invariant quantity), both forms are quite similar.}
\ba
e\vec{J}(\vec{r})=+2\mi0\vcE(\vec{r}) \, . \label{42}
\ea
Employing this relation into the first equation of (\ref{35}), the
sectors of $\vB$ and $\vcE$ fields become consistent.\ In other
words, the {\em ansatz} (\ref{42}) cuts away the $\vB'$ part of $\vB$
.\ Physically, what seems to happen is that
the $\vcE$-field (or more precisely, the matter background current,
$\vec{J}$) damps the effect of $\vB'$, at least, as the total
field felt by the electric charge.\\
\begin{sloppypar}
Returning to the presence of the fermionic current in eqs. (\ref{35}-\ref{38}), we shall interpret this current as a material background onto which the magnetic monopole
configurations show up.\ It is just in this sense that we distinguish between them and those of Dirac's types: these latter are classical configurations in the vacuum (Classical Electrodynamics in vacuum, to be more precise), and so, they need no material media for their {\em `existence'}.\ Even though, our monopoles {\em cannot appear} in vacua, they would configurate, for
exemple, in a superconductor medium, inside which the Cooper's
pairs of electrons would be this background (at any stationary
limit, because $e\nabla\cdot\vec{J}= 2\mi0 \nabla \cdot\vcE=0$).\ In addition, notice a similarity: both, the CSKR-model and a superconductor medium appear to have massive `photon'. \\ \end{sloppypar}

Another point that should be stressed concerns the background: we suppose -and this seems reasonable- that the charges acting as the sources for the electric and magnetic fields that yield the monopole configuration weakly affect the background, so that the back-reaction on the latter does not influence the conditions that allow {\em `monopole formation'}.\ However, if the density of charges becomes very high and the energy of the system of electric and magnetic fields is comparable to the energy density of the background, then our assumptions would be jeopardized.\ In short, we understand that we are relying on the approximation that the sources do not affect the background.\\
This background current seems to be very formal, introduced only to accomodate our monopole-like solution.\ The interesting question that now we raise is how to systematically propose a potential in the Dirac's equation in such a way that its solution, $\psi$, leads to a current $\vec{J}$ such that (\ref{42}) is fulfilled.\ From our analysis, we have obtained that an arbitrary potential, $V$, yielding a current given by (\ref{42}) does not lead to a separable form of the Dirac's equation.\ Imposing that $\vec{J}$ is known, $V$ is not uniquely fixed, i.e., different families of non-separable $V$ lead to the same expression for $\vec{J}$, and we are attempting at an explicit solution for $\psi$ as a result from the Dirac's equation with a particular potential.\\

Now, writing the non-relativistic Lagrangean for the system:
$L_p=\frac12 m\dot{r}^2 +q\vec{A}\cdot\dot{\vec{r}}$, with
$\vec{A}=\vec{A}^D$, and search for a conserved angular momentum
vector, we find\footnote{The first term is the angular momentum of a point-like object with momentum $\vec{p}$ and the second one comes from the interaction between the electromagnetic fields of both particles, the Poincar\'e' term.\ In addition, we know that in the quantum mechanical context its counterpart operator must commute with the Hamiltonian operator and satisfy the $SU(2)$ algebra.} $$
\vec{\cal J}=\vec{r}\wedge\vec{p} -\frac{gq}{4\pi c}\hat{r},$$and by quantizing its radial component (here, treated as a quantum
operator) according to Quantum Mechanics \cite{LWP}, we get: 
\ba
\hat{\bf r}\cdot{\bf {\cal J}}=\frac{n}{2} \quad \Longrightarrow
\quad \frac{qg}{4\pi\hbar c}=\frac{n}{2} \qquad n=1,2,\ldots.
\label{47}
\ea
Therefore, we obtain a {\em quantization condition} for the problem
(analogous to eq. (\ref{1})).\ [However, the difference put between the two types of such Abelian monopoles must
be remembered and taken into account].\ By using other procedures (e.g.
single-valuedness of the wave-function or Wu-Yang's approach) we are led to the same result, eq. (\ref{47}).\\

To close this section, we draw the attention to the fact that a
similar treatment to Proca's theory would lead us to a quite
analogous conclusion: this theory is compatible with the monopoles that were here
introduced.\ On the other hand, we justify our choice by
CSKR model because it presents another very interesting feature:
the mass parameter appears in a more general quantization
condition.\ This will be the goal of the following section.

\begin{sloppypar}\section{The `non-minimal' coupling and mass quantization}\end{sloppypar} 
In this section we shall introduce a new kind of `coupling' into
the model.\ This will be done by the following gauge covariant
derivative: $\nabla_\mu \psi(x)\equiv (\partial_\mu +\imath e A_\mu -\imath\sigma\tilde{G}_\mu)\psi(x)$, where $\sigma$ is the parameter
that measures the strength of the coupling between the fermions and
the tensorial sector.\ Hence, the model reads:
\ba {\cal L}_3 ={\cal L}_1 +\overline{\psi}\,(i\nabla_\mu\gamma^\mu -m_f)\psi\, ,\label{49}
\ea
(here, we choose $e,\sigma>0$, as we have already taken for $\mi0$)
.\footnote{A question must be asked: why the fermions are coupled to
$\tilde{G}_\mu$  and not to the gauge field $H_{\mu\nu}$ (as was done
for $A_\mu$)? We answer this question by saying that this is the
simplest form to write such ``coupling'' in a Lorentz-covariant way
and, at the same time, preserving the gauge invariance of the model.\ Nevertheless, it is clear that this vertex is non-renormalizable.\ Here, such aspect brings no major problems, since we are
dealing with a  non-relativistic Quantum Mechanical treatment.\ Actually, another ``coupling'' allowed in this way is: $\overline{\psi}\tilde{G}_\mu\gamma^\mu\gamma_5\psi$, what is clearly non-parity invariant; since we are not dealing with
aspects of parity breaking, we return to our former choice.}\\The influence of  non-minimal coupling on the 3-dimensional Maxwell-Chern-Simons model has been discussed in a series of works (some of them are listed in Ref. \cite{nmc3d}).

From (\ref{49}), there follow the dynamical eqs. for the femions:
\ba
& &  \left[(\imath\partial_\mu -eA_\mu +\sigma\tilde{G}_\mu)\gamma^\mu
-m_f \right]\psi(x)=0 \nn\\ & & \overline{\psi}(x)\left[(\imath\overleftarrow{\partial}_\mu+
eA-\mu -\sigma\tilde{G}_\mu)\gamma^\mu +m_f \right]=0 \nn \, ,
\ea
and those for the gauge fields:
\ba
& & \partial_\mu F^{\mu\nu}=-\mi0\epsilon^ {\nu\kappa\alpha\beta}
\partial_\kappa H_{\alpha\beta} +eJ^\nu = -2\mi0 \tilde{G}^\nu +e
J^\nu \, , \label{52.a} \\
& & \partial_\mu G^{\mu\nu\kappa}= +\epsilon^{\nu\kappa\alpha\beta}
\partial_\alpha \left(\mi0 A_\beta +\frac{\sigma}{2}J_\beta\right)
=+\mi0\tilde{F}^{\nu\kappa} +\frac{\sigma}{2}\epsilon^{\nu\kappa
\alpha\beta}\partial_\alpha J_\beta \, , \label{52.b}
\ea
and the Bianchi' identities (\ref{8}) as well.

Doing the same considerations as before to introduce magnetic
monopoles (static and point-like classical configuration onto
a matter background), we get the following equations:
\ba
& & \nabla\wedge\vB(\vec{r})= + e \vec{J}(\vec{r}) -2\mi0\vec{\cal E}
(\vec{r}) ,\quad \nabla\cdot\vE(\vec{r})=
 + e\rho(\vec{r}) -2\mi0{\cal B}(\vec{r})\label{53}, \\
& & \nabla\wedge\vec{\cal E}(\vec{r})=+\mi0\vB(\vec{r}) +\frac
{\sigma}{2} \nabla\wedge\vec{J}(\vec{r}) ,
\qquad \nabla \cdot \vec{\cal E}(\vec{r})=0 \\
& & \nabla \cdot \vB(\vec{r})= \chi^{0}(\vec{r})= g\delta^{3}(\vec{r})
 ,\qquad\nabla \wedge \vE(\vec{r})=0 , \\
& & \nabla{\cal B}(\vec{r})= -\mi0\vE(\vec{r}) +\frac{\sigma}{2} \nabla
\rho(\vec{r}) .\label{56} 
\ea
Now, we see that both sets of equations (one mixing $\cB$ and $\vE$ and another relating $\vB$ to $\vcE$) present inconsistencies.\ Fortunately, what happens here is that the 4-dimensional {\em ansatz}, eq. (\ref{41}), can solve all these problems\ Therefore, by implementing it in the above equations, we get (after ordering the equations):
\ba
& &\hspace{-1.4cm}\nabla\wedge\vB(\vec{r})= + e \vec{J}(\vec{r}) -2\mi0\vec{\cal E}(\vec{r})=0  ,\quad\nabla\cdot\vE(\vec{r})= + e\rho
(\vec{r}) -2\mi0{\cal B}(\vec{r})=0\label{57} \\
& & \nabla\wedge\vec{\cal E}(\vec{r})=\left( \frac{e\mi0}{e- \sigma
\mi0} \right) \vB(\vec{r}),\qquad\nabla \cdot
\vec{\cal E}(\vec{r})=0  \label{58}, \\
& & \nabla \cdot \vB(\vec{r})= + g\delta^{3}(\vec{r}) ,
\quad\qquad\qquad\nabla \wedge \vE(\vec{r})=0  , \\
& & \nabla{\cal B}(\vec{r})= - \left( \frac{e\mi0}{e-\sigma \mi0} \right)
\vE(\vec{r}). \label{61} 
\ea
It is clear that, hereafter we shall be considering regimes of the model for which $e\neq\sigma\mi0$ is satisfied.

Now, placing a particle (with electric $q=eq'$ and ``tensorial''
$Q=\sigma$ charges\footnote{The ``current equation'' for the tensorial
sector may be writen as: $$\partial_\mu G^{\mu\nu\kappa}=j^{\nu\kappa}
\quad {\rm{\mbox with}}\qquad j^{\mu\nu}=\left\{\begin{array}{l}j^{0i}
\equiv (\vec{j}_1)^i \\ j^{ij}\equiv \epsilon^{ijk}(\vec{j}_2)^k 
\end{array} \right. \, . $$(it's clear that the conservation equation
for $j_{\mu\nu}$ reads $\partial_\mu j^{\mu\nu}=0$).\ From this, we see
that this sector carries no charge attribute.\ What happens is that all
fermions carry the same {\em charge} with respect to the tensorial gauge group,
$Q=\sigma$.}; mass $m$) into the external magnetic field (let us recall the
assumptions from Section 2, concerning matter background), we get
its non-relativistic Lagrangian:
$$L_2=\frac12 m\dot{r}^2 +q\vA\cdot\dot{\vec{r}} -\sigma\vcE\cdot\dot
{\vec{r}}.$$
And, the conserved angular momentum vector reads: $$\vec{\cal J}=\vec{r}\wedge\vec{p} -\frac{g}{4\pi c} \left(q+\frac{e\sigma\mi0}{e-\sigma\mi0}\right)\hat{r}.$$
Now, the second term, that is related with the {\em `electromagnetic'} angular momentum, brings us information about the tensorial gauge sector, by defining an
{\em `effective charge'} as: $\left(q+\frac{e\sigma\mi0}{e-\sigma\mi0}
\right)$.\\

Now, in the context of Quantum Mechanics, we quantise the radial
component of the conserved angular momentum operator:
\ba
\hat{\bf r}\cdot{\bf{\cal J}}=\frac{n}{2} \quad\Longrightarrow\quad
\frac{g}{4\pi\hbar c}\left(q+\frac{e\sigma\mi0}{e-\sigma\mi0}\right)=
\frac{n}{2} \, ,\label{64}
\ea
(with $n$ integer).\ Hence, we obtain what we have announced: the presence of the mass parameter into a generalized quantization condition.\ Two limits
of the relation above take importance:
\ba
\lim_{\sigma\to 0}\quad \longmapsto \frac{qg}{4\pi\hbar c}=\frac{n}{2}
\qquad{\rm{\mbox  and}}\qquad  \lim_{\mi0\to 0}\quad \longmapsto
\frac{qg}{4\pi\hbar c}=\frac{n}{2}\label{65}  
\ea
\begin{sloppypar}
From (\ref{65}), we see that, at the limit $\sigma\to 0$ we recover the
result obtained in Section 2, which is expected.\ But, if we take
$\mi0\to 0$ we recover the same result.\ This seems to say that the
interaction between the fermions and the tensorial sector is performed
by means of the topological term, that linkes both gauge symmetries.\ Such a issue should be better sought.\\
\end{sloppypar}
It is worth noticing that the topological mass parameter does not get shifted by 1\_loop corrections induced by loop of matter fields (scalars and/or spinors) minimally coupled to $A_{\mu}$, but non-minimally coupled to $H_{\mu\nu}$.\ Indeed, by computing the self-energy diagram that exhibits $A_{\mu}$ to $H_{\nu\rho}$ on the external legs, it has been shown that the (finite) fermionic 1\_loop contribution does not shift the mass parameter $\mi0$, so that the quantization condition displayed in (\ref{64}) does not suffer from (finite) renormalization effects on $\mi0$.\ Such a Feynman graph and its answer (for the case of scalar matter fields) are quoted in the Appendix.\\
\section*{Concluding Remarks}
The main motivation of the present paper was the investigation of the possibility for the existence of Dirac's monopole solutions associated to the massive spin\_1 model described by the mixing of a vector gauge field to a rank-2 tensor gauge field according to CSKR.\ We have concluded that no such a monopole emerges if matter is absent.\ Indeed, we have been able to work out possible conditions on the matter background so as to trigger monopole formation.\ We would however like to understand better the r\^ole of the matter background on the physics of the monopole.\ For example, quantization of the latter in the presence of the background; or still, possible bounds on the monopole mass as dictaded by the background.\\

Our quantization relation involving the topological mass parameter does not mean that the latter is quantized as it is the case for the topological mass parameter in Abelian \cite{HT} or non-Abelian \cite{NACS} Chern-Simons theory in (2+1) dimensions .\ All we get here is a quantization condition where all the parameters are mixed.\ If we assume electric (as well as magnetic) charge quantization, then we get the quantization of the product $\sigma\mi0$.\ However, this quantization condition should be more deeply exploited. \\
Moreover, at the attempt of take some light to our motivating question, we have noticed that the non-coexistence of massive vector boson and Dirac's monopole might lie in the way Electrodynamics-type models are built up, i.e., in terms of 2-form field-strength (containing the classical physical fields); and not in the way of mass generation, as we had initially suspected.\ [Indeed, in (2+1)D Maxwell-Chern-Simons theory -describing a massive spin$\_$1 boson- presents similar trouble concerning the introduction of Dirac-like monopole, see Ref. \cite{HT} for details.]\vspace{.4cm} \\
\centerline{\bf Acknowledgments}
\vspace{0.1cm}\\
The authors would like to thank Dr. R. Casana Sifuentes (CBPF/DCP) for his assistance concerning computer facilities, Dr. A.L.M.A. Nogueira (CBPF/DCP) for drawing our attention to Ref.[22] and CNPq-Brasil for the financial support.\\
\appendix
\section*{\centerline{ Appendix}}\vspace{.1cm}
The Feynman graph that exhibits $A^{\mu}$ and $H^{\nu\kappa}$ on the external legs with a loop of scalars\footnote{A similar graph with spinor loop (instead of scalar one) may lead us to a slightly different result, but no shift of the mass parameter will occur by finite 1\_loop contributions.} is depicted below:\vspace{.5cm}
\ba
\setlength{\unitlength}{1mm}
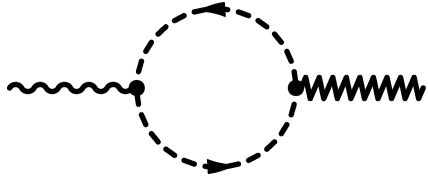
{\begin{fmffile}{loop11}
\parbox{60mm}{\begin{fmfgraph*}(55,25) \nn \fmfpen{thick}
\fmfleft{i1}  \fmfright{o1} 
\fmf{photon,label=$A_{\mu}$,label.side=left}{i1,v1}  
\fmf{scalar,right,tension=0.4,label=$k+p$}{v1,v2} 
\fmf{scalar,right,tension=0.4,label=$k$}{v2,v1}
\fmf{zigzag,label=$H^{\beta\lambda}$,label.side=left}{v2,o1}  \fmfdotn{v}{2} 
\end{fmfgraph*}}
\end{fmffile}}\nn\\
\ea
The result of the above graph, after dimensional regularisation has been adopted, reads as follows:
\ba
I_{\mu}^{\hspace{0.07in}\beta\lambda}(p)& =&-i\pi^2\epsilon^{\nu\alpha\beta\lambda}\,p_{\alpha}\left\{p_{\mu}p_{\nu}\int_{0}^{1}dz\, (1-2z)^2 \ln[p^2 z(1-z) -m^2]+\right. \nn \\ 
& & \left. -\frac13\left(\frac{2}{\delta} -k\right) p_\mu p_\nu -2\eta_{\mu\nu}\left[\left(\frac{2}{\delta}-k+1\right)\left(m^2-\frac{p^2}{6}\right)\right.\right.\nn\\
& & \left.\left.+\int_0^1 dz\, (p^2 z(1-z) -m^2)\ln[p^2 z(1-z) -m^2]\right]\right\},\nn
\ea
here, $\delta$ ($=4-D$) is the dimensional regularisation parameter and $k\equiv\gamma+\ln\pi$ ($\gamma$ is the Euler's constant); $m^2=2\mi0^2$ is the mass parameter.\\The finiteness of these integrals (written in terms of Feynman parametrisation) is evident.


\begin{thebibliography}{99}
\bibitem{Dirac1} P.A.M. Dirac, Proc. Roy. Soc. \underline{A133}
(1931)60; Phys. Rev. \underline{74}(1948)817;
\bibitem{GOlive} P. Goddard and D. Olive, Rep. Prog. Phys. 
\underline{41}(1978)1357;
\bibitem{Giamb} C. Bollini and J.J. Giambiagi, Nucl. Phys. \underline{B123} (1977)311;
\bibitem{tHooft}G. 't Hooft, Nucl. Phys. \underline{B79} (1974)276;
\bibitem{Polyakov}A.M. Polyakov, JETP Lett. \underline{20}(1974)194;
\bibitem{GG}H. Georgi and S.L. Glashow, Phys. Rev. \underline{D6} (1972)2977;
\bibitem{extension}C.N. Yang, Phys. Rev. \underline{D1} (1970)2360; A.S. Schwarz, Nucl. Phys. \underline{B112} (1976)358; D. Olive, Nucl. Phys. \underline{B113} (1976)413;
\bibitem{SW}N. Seiberg and E. Witten, Nucl. Phys. \underline{B426} (1994)19; {\em ibd} \underline{B430} (1994)485, ERRATUM;
\bibitem{Giacomelli}C. Giacomelli, in ``Theory and Detection of Magnetic Monopoles in Gauge Theories'', ed. N. Craigie (World Scientific, 1986) Sections 7.1, 7.2 and 7.6;
\bibitem{YJ}A.Yu. Ygnatiev and G.C. Joshi, Phys. Rev. \underline
{D53} (1996)984;
\bibitem{Singleton}D. Singleton, Int. J. Th. Phys. \underline{35} (1996)2419;
\bibitem{Ahrens}T. Ahrens, Nuovo Cim.  \underline{103A} (1990)1139;
\bibitem{Dattoli}G. Dattoli {\em et al}, Lett. Nuovo Cim. \underline{20} (1977)686;
\bibitem{CS}C. Cremmer and J. Scherk, Nucl. Phys. \underline{B72}
(1974)117;
\bibitem{KR}M. Kalb and P. Ramond, Phys. Rev. \underline{D9}
(1974)2273;
\bibitem{ABL}T.D. Allen, M.J. Bowick and A. Lahiri, Mod. Phys. Lett 
\underline{A6}(1991)559;
\bibitem{ABN}R. Amorim and J. Barcelos-Neto,  Mod. Phys. Lett 
\underline{A10}(1995)917; 
\bibitem{tese} Winder A. Moura-Melo, M.Sc. Thesis (CBPF, 1997)[portuguese];
\bibitem{cax97} W.A. Moura-Melo, N. Panza and J.A. Helay\"el-Neto, in ``Proc. XVIII Brazilian Nat. Meet. Particles and Fields'' (Soc. Bras. F\'{\i}sica, 1997)146;
\bibitem{WuYang}T.T. Wu and C.N. Yang, Phys. Rev. \underline{D12} (1975)3845;
\bibitem{LWP}H.J. Lipkin, W.I. Weisberger and M. Peshkin, Ann. Phys. \underline{53} (1969)203;
\bibitem{nmc3d}S.K. Paul and A. Khare, Phys. Lett. \underline{B193} (1987)253; I. Kogan, Phys. Lett. \underline{B262} (1991)83; M. Torres, Phys. Rev. \underline{D46} (1992)R2295; H.Cristiansen, M. Cunha, J.A. Helay\"el-Neto, L. Mansur and A.L.M.A. Nogueira, Int. J. Mod. Phys. \underline{A14} (1999)147;
\bibitem{HT}M. Henneaux and C. Teitelboim, Phys. Rev. Lett. \underline{56} (1986)689;
\bibitem{NACS}S.Deser, R. Jackiw and S. Templeton, Ann. Phys. \underline{140} (1982)372; {\em ibd} Phys. Rev. Lett. \underline{48} (1982)975.
\end{thebibliography}
\end{document}